\documentclass{article}

    \PassOptionsToPackage{numbers, compress}{natbib}

    \usepackage[preprint]{neurips_2019}

\usepackage[utf8]{inputenc} 
\usepackage[T1]{fontenc}    
\usepackage{hyperref}       
\usepackage{url}            
\usepackage{booktabs}       
\usepackage{amsfonts}       
\usepackage{nicefrac}       
\usepackage{microtype} 
\usepackage{graphicx}
\usepackage{textcomp}
\usepackage{xcolor}

\title{State-of-the-art Speech Recognition using EEG and Towards Decoding of Speech Spectrum From EEG}

\author{%
  Gautam Krishna\thanks{Equal author contribution} \\
  Brain Machine Interface Lab\\
  The University of Texas at Austin\\
  \texttt{} \\
   \And
   Yan Han\thanks{Equal author contribution} \\
   Brain Machine Interface Lab \\
   The University of Texas at Austin \\
   \AND
   Co Tran\thanks{Equal author contribution} \\
   Brain Machine Interface Lab \\
   The University of Texas at Austin \\
   \And
   Mason Carnahan \\
    Brain Machine Interface Lab \\
   The University of Texas at Austin \\
   \And
   Ahmed H Tewfik \\
   Brain Machine Interface Lab  \\
    The University of Texas at Austin \\
}

\begin{document}

\maketitle

\begin{abstract}
  In this paper we first demonstrate continuous noisy speech recognition using electroencephalography (EEG) signals on English vocabulary using different types of state of the art end-to-end automatic speech recognition (ASR) models, we further provide results obtained using EEG data recorded under different experimental conditions. We finally demonstrate decoding of speech spectrum from EEG signals using a long short term memory (LSTM) based regression model and Generative Adversarial Network (GAN) based model. Our results demonstrate the feasibility of using EEG signals for continuous noisy speech recognition under different experimental conditions and we provide preliminary results for synthesis of speech from EEG features.
  
\end{abstract}

\section{Introduction}

Electroencephalography (EEG) is a non-invasive way of measuring electrical activity of human brain. In \cite{krishna2019speech} authors demonstrated deep learning based automatic speech recognition (ASR) using EEG signals for a limited English vocabulary of four words and five vowels. In \cite{krishna20} authors demonstrated continuous ASR using the same set of EEG features used in \cite{krishna2019speech} for larger English vocabulary.  The work presented in this paper is different from work presented in reference \cite{krishna20} as this paper introduces two new sets of EEG features, provide EEG based speech recognition results for additional conditions like listen, listen and spoken. In \cite{krishna20} authors provided results only for spoken condition. In addition, in this paper we provide speech recognition results using a new end-to-end model called RNN transducer model and also demonstrate preliminary results for speech synthesis using EEG signals. Finally in this paper we provide speech recognition results evaluated on data sets consisting of more number of subjects than the ones used in reference \cite{krishna20}. 

Recently in \cite{anumanchipalli2019speech} researchers demonstrated synthesizing speech from electrocorticography (ECoG) signals recorded for spoken English sentences. ECoG is an invasive technique for measuring electrical activity of human brain. In \cite{ramsey2017decoding} authors demonstrated speech recognition using ECoG signals. In \cite{zhao2015classifying} the authors used classification approach for identifying phonological categories in imagined and silent speech. In this paper we demonstrate continuous noisy speech recognition using EEG signals recorded in parallel with speech for spoken English sentences, EEG signals recorded in parallel while the subjects were listening to utterances of the same English sentences and finally we demonstrate speech recognition by concatenating both this sets of EEG features. Inspired from the unique robustness to environmental artifacts exhibited by the human auditory cortex \cite{yang1991auditory,mesgarani2011speech} we used EEG data recorded in presence of background noise for this work and demonstrated lower word error rate (WER) for smaller corpus using EEG features. We first conducted speech recognition experiments using the EEG features used by authors in \cite{krishna2019speech,krishna20} and we further conducted experiments using two more different feature sets which are more commonly used by neuroscientists studying EEG signals. In this paper we provide comparison of the speech recognition performance results obtained using these different feature sets. 

EEG has the big advantage of being a non invasive technique compared to ECoG which is an invasive technique, making EEG based brain computer interface (BCI) technology easily deployable and it can be used by subjects without the need of undergoing a neurosurgery to implant ECoG electrodes.  We believe speech recognition using EEG will help people with speaking disabilities to use voice activated technologies with better user experience, help with speech restoration and also potentially introduce a new form of thought based communication.  

Inspired from the results presented in \cite{anumanchipalli2019speech} we used long short memory (LSTM) \cite{hochreiter1997long} based regression model, generative adversarial network (GAN) \cite{goodfellow2014generative}, wasserstein generative adversarial networks (WGAN) \cite{arjovsky2017wasserstein} to decode the Mel-frequency cepstral coefficients (MFCC) features of the audio that the subjects were listening from the EEG signals which were recorded in parallel while they were listening to the audio as well as we decode MFCC features of the sound that the subjects spoke out from the EEG signals which were recorded in parallel with their speech.

\section{Automatic Speech Recognition System Models}
In this section we briefly describe the ASR models that were used in this work. We used end to end ASR models which directly maps the EEG features to text. We did experiments using three different types of end to end ASR models, namely: Connectionist Temporal Classification (CTC) model \cite{graves2006connectionist,graves2014towards}, Attention based RNN encoder decoder model \cite{cho2014learning,chorowski2015attention,bahdanau2014neural} and RNN transducer model \cite{graves2012sequence,graves2013speech}. For all the models the number of time steps of the encoder was equal to the product of sampling frequency of EEG features and sequence length. Since different subjects spoke with different rate and listening utterances were of different length, there was no fixed value for the encoder time steps, so we used Tensorflow's dynamic RNN cell for the encoder.  

\subsection{Connectionist Temporal Classification (CTC)}

In our work we used a single layer gated recurrent unit (GRU) \cite{chung2014empirical} with 128 hidden units as encoder for the CTC network. The decoder consists of a combination of a dense layer and a softmax activation. The output at every time step of the GRU layer is fed into the decoder network. 
We used CTC loss function with adam optimizer \cite{kingma2014adam} and during inference time we used CTC beam search decoder. The mathematical details of CTC loss function computation is covered in \cite{graves2014towards,krishna20}. 

A dynamic algorithm is used to compute the CTC loss.
In our work we used character based CTC ASR model and the model was trained for 800 epochs to observe loss convergence. 

\subsection{RNN Encoder-Decoder or Attention model}

RNN encoder - decoder ASR model consists of a RNN encoder and a RNN decoder with attention mechanism. We used a single layer GRU with 512 hidden units for both encoder and decoder. A dense layer followed by softmax activation is used after the decoder GRU to get the prediction probabilities. We used cross entropy as loss function with adam as the optimizer. We used teacher forcing algorithm \cite{williams1989learning} to train the model. The model was trained for 150 epochs to observe loss convergence. 
During inference time we used beam search decoder. The labels are augmented using two special tokens namely the start token and end token which indicates beginning and end of a sentence. During inference time the label prediction process stops when the end token label is predicted.

The mathematical details of the attention mechanism used in our attention model are covered in references \cite{krishna20,bahdanau2014neural,chorowski2015attention}. More specifically we used the exact attention mechanism used by authors in \cite{krishna20}. 
\subsection{RNN Transducer model}

The RNN transducer model consists of an encoder model working in parallel with a prediction network over the output tokens. We used LSTM with 128 hidden units for both our encoder and prediction network. The encoder and prediction network outputs are passed to a joint network which uses tanh activation to compute logits, which are passed to softmax layer to get the prediction probabilities. During inference time, beam search decoder was used. The RNN transducer model was trained for 200 epochs using stochastic gradient descent optimizer to optimize RNN T loss\cite{graves2012sequence}. We used character based RNN transducer model for this work. More details of RNN transducer model are covered in \cite{graves2012sequence,graves2013speech}.

\section{Design of Experiments for building the database}
We built three databases for this work. All the subjects who took part in the experiments were healthy UT Austin undergraduate, graduate students in their early twenties for all the databases. For the first database A, 20 subjects took part in the experiment. Out of the 20 subjects, 8 were females and rest were males. Only five out of the 20 subjects were native English speakers. Each one of them was asked to speak the first 9 English sentences from USC-TIMIT database\cite{narayanan2014real} three times and their simultaneous speech and EEG signals were recorded. The sentences were shown to them on a computer screen. This data was recorded in presence of background noise of 65dB. Music played from our lab computer was used as the source of generating background noise.    

For the second database B, 15 subjects took part in the experiment. Out of the 15 subjects, three were females and rest were males. Only two out of the 15 subjects were native English speakers. Each one of them was asked to listen to the utterances of the first 9 English sentences from USC-TIMIT database\cite{narayanan2014real} and then they were asked to speak the utterances that they listened to. Their EEG was recorded in parallel while they were listening to the utterances and also their simultaneous speech and EEG signals were recorded while they were speaking out the utterances that they listened to. This data was recorded in presence of background noise of 50dB. The utterances that the subjects listened to were also recorded. Then the 15 subjects were asked to repeat the same experiment two more times. 

For the third database C, five female and five male subjects took part in the experiment. Each one of them was asked to read out the first 30 sentences from USC-TIMIT database \cite{narayanan2014real} and their simultaneous speech and EEG signals were recorded. This data was recorded in absence of external background noise. Then the 10 subjects were asked to repeat the same experiment two more times.

Throughout this paper we will refer to the acoustic features for spoken speech as spoken MFCC, acoustic features for the listening utterances that were recorded as listen MFCC, EEG features recorded in parallel with spoken speech as spoken EEG and EEG features recorded in parallel while the subjects were listening to the utterances as listen EEG.

We used Brain Vision EEG recording hardware. Our EEG cap had 32 wet EEG electrodes including one electrode as ground. We used EEGLab \cite{delorme2004eeglab} to obtain the EEG sensor location mapping. It is based on standard 10-20 EEG sensor placement method for 32 electrodes.

\section{EEG and Speech feature extraction details}

For preprocessing of EEG we followed the same method as described by the authors in \cite{krishna2019speech,krishna20}. EEG signals were sampled at 1000Hz and a fourth order IIR band pass filter with cut off frequencies 0.1Hz and 70Hz was applied. A notch filter with cut off frequency 60 Hz was used to remove the power line noise. EEGlab's \cite{delorme2004eeglab} Independent component analysis (ICA) toolbox was used to remove other biological signal artifacts like electrocardiography (ECG), electromyography (EMG), electrooculography (EOG) etc from the EEG signals. We extracted three different EEG feature sets for this work. All EEG features were extracted at a sampling frequency of 100 Hz. The first EEG feature set was same as the ones used by the authors in \cite{krishna2019speech,krishna20} namely root mean square, zero crossing rate, moving window average, kurtosis and power spectral entropy with frequency bands value equal to none (power per band was same as the power spectral density). For this set, EEG feature dimension was 31(channels) $\times$ 5 or 155. The second set of features were the magnitudes of short time Fourier Transform of the EEG signals, discrete time wavelet based spectral entropy using approximation and detailed coefficients. We used db4 wavelet. Frequency bands value was kept as none. For every window we extracted only level one coefficients. For this set, EEG feature dimension was 31(channels) $\times$ 3 or 93. The third set of features were power spectral entropy based on delta, theta, alpha and beta EEG frequency bands \{0.5,4,7,12,30\}Hz, hurst exponent and petrosian fractal dimension. This features are more commonly used by neuroscientists studying EEG signals. For this set, EEG feature dimension was 31(channels) $\times$ 3 or 93. For both listening speech and spoken speech we extracted MFCC 13 features and then we computed first and second order differentials (delta and delta-delta) thus having total MFCC 39 features. The MFCC features were also sampled at 100Hz frequency. For spectral entropy calculations, we used the python neurokit library. For hurst exponent and petrosian fractal dimension calculation we used python pyeeg library. 

\section{EEG Feature Dimension Reduction Algorithm Details}

We used non linear dimension reduction methods to denoise the EEG feature space. The tool we used for this purpose was Kernel Principle Component Analysis (KPCA) \cite{mika1999kernel}. We plotted cumulative explained variance versus number of components to identify the right feature dimension for each feature set. We used KPCA with polynomial kernel of degree 3 \cite{krishna2019speech,krishna20}. For the first feature set, 155 dimension was reduced to 30. For the second feature set, 93 dimension was reduced to 50. For the third feature set when we plotted explained variance versus number of components we observed that it was best to keep the original dimension. 
We used python scikit library for performing KPCA. The cumulative explained variance plot is not supported by the library for KPCA as KPCA projects features to different feature space, hence for getting explained variance plot we used normal PCA but after identifying the right dimension we used KPCA to perform dimension reductions.  

We further computed delta, delta and delta features, thus the final feature dimension of EEG feature set 1 was 90 (30 $\times$ 3), for feature set 2 final dimension was 150 (50 $\times$ 3) and for feature set 3 final dimension was 279 (93 $\times$ 3). The same preprocessing and dimension reduction methodology was followed for both spoken EEG and listen EEG data. 

\section{Models to predict Listen MFCC from Listen EEG}

In this section we briefly describe the architectures of the deep learning models that we used to predict listen MFCC features from listen EEG features. For this decoding problem we considered feature set 1, feature set 2 and feature set 3 for listen EEG and delta, delta-delta features were not considered, hence the listen MFCC dimension was 13. We tried two different approaches to solve this problem. 1) using a LSTM based regression model and 2) using generative model. For both the approaches, during test time, we used three evaluation metrics: RMSE, Normalized RMSE and Mel cepstral distortion (MCD)between the model output and listen MFCC features from test set. The RMSE values were normalized by dividing the RMSE values with the absolute difference between the maximum and minimum value in the test set observation vector. 

\subsection{LSTM based regression model}

Our LSTM based regression model consists of two layers of LSTM with 128 hidden units in each layer. The final LSTM layer is connected to a time distributed dense layer with 13 hidden units. Root mean squared error was used as the loss function and the model was trained for 200 epochs to observe loss convergence and adam optimizer was used \cite{kingma2014adam}. 90 \% of the data was used to train the model and remaining 10 \% was used as test set.  

\subsection{Generative Model}

Generative Adversarial Network (GAN) consists of two networks namely the generator model and the discriminator model which are trained simultaneously. The generator model learns to generate data from a latent space and the discriminator model evaluates whether the data generated by the generator is fake or is from true data distribution. The training objective of the generator is to fool the discriminator. 

\begin{figure}[h]
\centering
\includegraphics[height=5cm, width=0.4
\textwidth,trim={0.1cm 0.1cm 0.1cm 0.1cm},clip]{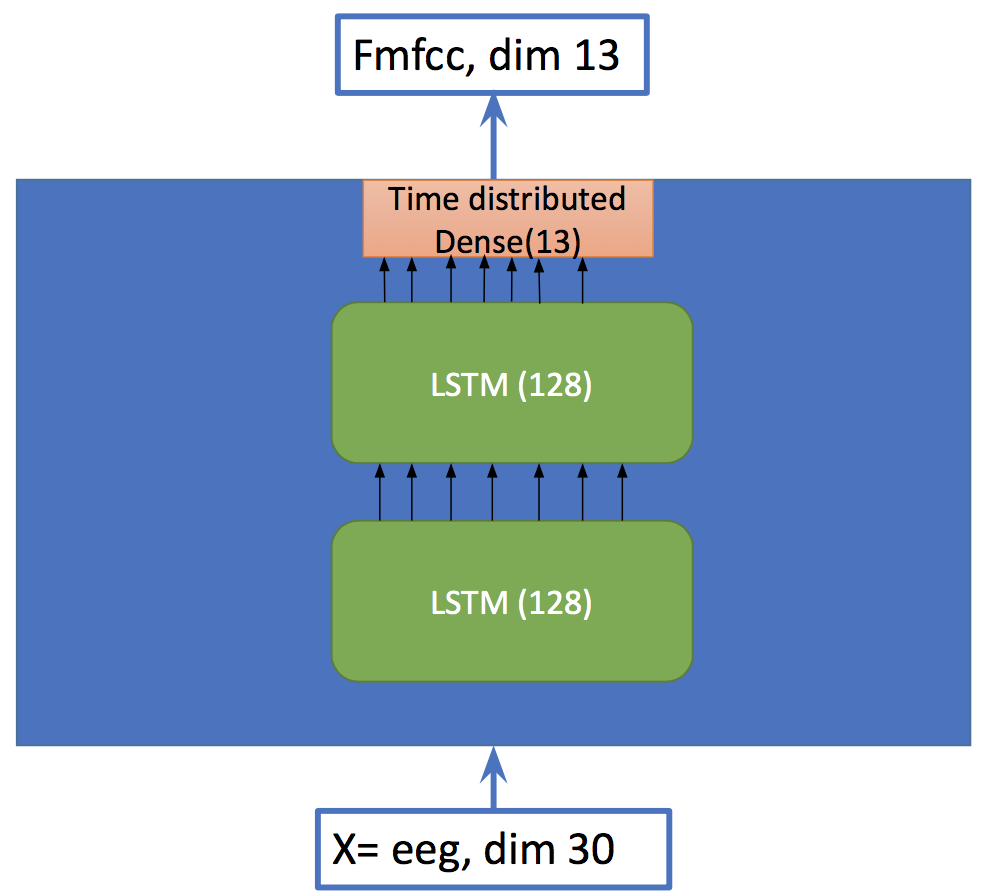}
\caption{Generator in our GAN model}
\label{1vsall}
\end{figure}

\begin{figure}[h]
\centering
\includegraphics[height=5cm, width=0.8
\textwidth,trim={0.1cm 0.1cm 0.1cm 0.1cm},clip]{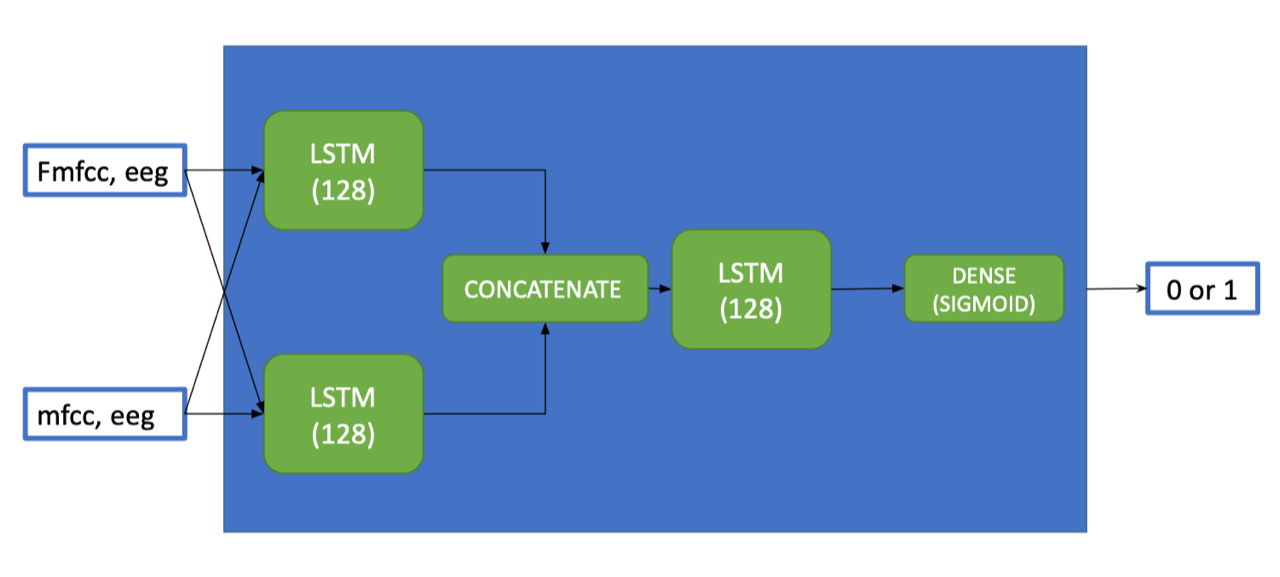}
\caption{Discriminator in our GAN model} 
\label{1vsall}
\end{figure}

Our generator model consists of two layers of LSTM with 128 hidden units in each layer followed by a time distributed dense layer with 13 hidden units. During training, real listen EEG features say with dimension 30 (when used with EEG feature set 1) from training set are fed into the generator model and the generator outputs a vector of dimension 13, which can be considered as fake listen MFCC. This is explained in figure 1. In the figure 1, Fmfcc denotes fake MFCC. The discriminator model consists of two single layered LSTM's with 128 hidden units connected in parallel. At each training step a pair of inputs are fed into the discriminator. The discriminator takes (real listen EEG features, fake listen MFCC) and (real listen EEG features, real listen MFCC) pairs.  The outputs of the two parallel LSTM's are concatenated and then fed to another LSTM with 128 hidden units. The last time step of the final LSTM is fed into the dense layer with sigmoid activation function. This is explained in figure 2. 

In order to define the loss functions for both our generator and discriminator model let us first define few terms. Let $P_{e_f}$ be the sigmoid output of the discriminator for (real listen EEG features, fake listen MFCC) input pair and let $P_{e_s}$ be the sigmoid output of the discriminator for (real listen EEG features, real listen MFCC) input pair. Then we can define the loss function of generator as $-\log (P_{e_f})$ and loss function of discriminator as $-\log (P_{e_s}) - \log(1-P_{e_f})$. In order to get better stabilized training, we also tried implementing the same idea using WGAN \cite{arjovsky2017wasserstein} where the loss function is earth mover's distance or wasserstein 1 distance instead of the log loss. Both GAN and WGAN models were trained for 500 epochs and adam optimizer was used \cite{kingma2014adam}. During test time listen EEG features from the test set is fed into the generator and generator generates listen MFCC features. 90 \% of the data was used as the training set and remaining as test set. 

\section{Predicting Spoken MFCC from Spoken EEG}
For predicting spoken MFCC from spoken EEG we used database C, same models used for predicting listen MFCC from listen EEG but we considered only feature set 1 for spoken EEG and delta, delta-delta features were not considered, hence the spoken MFCC dimension was 13 and spoken EEG dimension was 30 for this particular decoding problem. 

\section{Results}

The attention model was predicting a word at every time step while the other two types of ASR models were predicting a character at every time step. For CTC and RNN transducer, word based model training was not stable, hence we used only character based model for both CTC and RNN transducer. The performance metric for attention model was word error rate (WER) and character error rate (CER) was the performance metric for both CTC and RNN transducer model. For both attention model and CTC model for both data sets A and B, 80 \% of the data was used as training set, 10 \% for validation set and remaining 10 \% for test set. For RNN transducer model, for data set A data from the first 18 subjects was used as training set and remaining two subjects data as validation and test set respectively. For data set B, data from the first 12 subjects was used as training set, next two subjects data was used as validation set and last subject data was used as test set. We observed that this way of data splitting reduced the over fitting phenomena for all the models. We used Nvidia Quadro GV100 GPU with 32 GB video memory for training all the models. 

Table 1 shows the results obtained during test time for CTC and attention model when the models were trained using only data set A with feature set 1 EEG features. Table 2 shows the test time results for RNN transducer model for the same data set. Tables 3, 4, 5, 6 and 7 shows the results obtained during test time for CTC and attention model when the models were trained using data set B for different experiments.

In general we observed that for all the models that were used to perform speech recognition using EEG features, the error rate during test time went up as the corpus size increase. We believe as the corpus size increase the deep learning models need to be trained with more number of examples to achieve better performance during test time. We also observed that the models gave comparable performance when trained with the EEG feature set 1 and feature set 3. As the corpus increase, training with feature set 2 gave higher error rates compared to other two feature sets, hence we didn't perform more experiments with feature set 2. Table 5 shows the test time result for feature set 2 training with different models for one type of experiment. 
We didn't perform more experiments with the RNN transducer model as it demonstrated poor performance even when tested on smaller corpus for the different feature sets. 

In \cite{krishna2019speech} authors demonstrated that EEG sensors T7 and T8 contributed most to the ASR test time accuracy, hence we tried performing ASR experiments using EEG data from only T7, T8 electrodes without performing dimension reduction using data set B with EEG feature set 1 and we observed error rates of 70 \% WER for attention model for spoken EEG, 73.3 \% WER for attention model for listen EEG, \textbf{68} \% CER for CTC model for spoken EEG and \textbf{66} \% CER for CTC model for listen EEG during test time for predicting 9 sentences ( the complete test time corpus). We noticed that the CTC model error rate for listen EEG using T7, T8 sensors data was slightly lower than the error rate obtained when the model was trained using the data from all 31 sensors followed by dimension reduction as seen from table 3. The other error rates obtained after T7, T8 training were comparable (but were not lower) to the results obtained when the models were trained using the data from all 31 sensors followed by dimension reduction for other ASR experiments.  

For predicting listen MFCC features from listen EEG features, the LSTM regression model demonstrated lowest average normalized RMSE, RMSE and MCD during test time compared to GAN and WGAN models as shown in tables 8,9 and 10. WGAN model demonstrated better test time performance than GAN model when trained using EEG feature set 1 and 2. Also the WGAN model demonstrated better training loss convergence for both generator and discriminator models compared to the GAN's generator and discriminator models for all the EEG feature sets.  

For predicting spoken MFCC features from spoken EEG features, the LSTM regression model again demonstrated lowest average normalized RMSE, RMSE and MCD during test time compared to GAN and WGAN models as shown in table 11. 

We computed WER values for CTC and RNN transducer model for some experiments, in our opinion since CTC and RNN-T models were predicting characters at every time step, CER is a better performance metric than WER for those models. For data set B, for listen condition the CTC model gave WER values 52.6 \%, 87.09 \%, 88.88 \% and 94.9 \% for number of sentences = \{3,5,7,9\} respectively using EEG feature set 1 and for spoken condition the same model gave WER values 73.6 \%, 83.8 \%, 91.1\% and 91.5 \% respectively using the same EEG feature set 1.  

For data set B, for listen condition the RNN-T model gave WER values 92.98 \%, 69.89 \%, 70.37 \% and 92.66 \% for number of sentences = \{3,5,7,9\} respectively using EEG feature set 1 and for spoken condition the same model gave WER values 64.91 \%, 82.8 \%, 79.26\% and 93.79 \% respectively using  EEG feature set 2.

\begin{table}[!ht]

\caption{Results for data set A with feature set 1}

\centering
\begin{tabular}{lllll}
\hline
\textbf{\begin{tabular}[c]{@{}l@{}}Number\\ of \\ Sentences\end{tabular}} & \textbf{\begin{tabular}[c]{@{}l@{}}Number of \\ unique \\ characters\\ contained\end{tabular}} & \textbf{\begin{tabular}[c]{@{}l@{}}Number of\\ unique\\ words\\ contained\end{tabular}} & \textbf{\begin{tabular}[c]{@{}l@{}}CTC \\ Model:\\ Spoken\\ EEG\\ (CER \%)\end{tabular}} & \textbf{\begin{tabular}[c]{@{}l@{}}Attention\\  Model:\\ Spoken\\ EEG\\ (WER \%)\end{tabular}} \\ \hline
3                                                                         & {\color[HTML]{000000} 19}                                                                      & 19                                                                                      & 32.5                                                                                     & 0                                                                                              \\ \hline
5                                                                         & {\color[HTML]{000000} 20}                                                                      & 29                                                                                      & 54                                                                                       & 54.8                                                                                           \\ \hline
7                                                                         & 22                                                                                             & 42                                                                                      & 67.5                                                                                     & 64.4                                                                                           \\ \hline
9                                                                         & {\color[HTML]{000000} 23}                                                                      & 55                                                                                      & 69                                                                                       & 60                                                                                             \\ \hline
\end{tabular}
\end{table}

\begin{table}[!ht]

\caption{Results for RNN Transducer model on data set A with feature set 1}

\centering
\begin{tabular}{lll}
\hline
\textbf{\begin{tabular}[c]{@{}l@{}}Number\\ of \\ sentences\end{tabular}} & \textbf{\begin{tabular}[c]{@{}l@{}}Number of \\ unique \\ characters\\ contained\end{tabular}} & \textbf{\begin{tabular}[c]{@{}l@{}}Spoken\\ EEG\\ (CER\%)\end{tabular}} \\ \hline
3                                                                         & {\color[HTML]{000000} 19}                                                                      & 41.09                                                                   \\ \hline
5                                                                         & {\color[HTML]{000000} 20}                                                                      & 60.34                                                                   \\ \hline
7                                                                         & 22                                                                                             & 66.22                                                                   \\ \hline
9                                                                         & {\color[HTML]{000000} 23}                                                                      & 64.26                                                                   \\ \hline
\end{tabular}
\end{table}

\begin{table}[!ht]

\caption{Results for data set B with feature set 1}

\centering
\begin{tabular}{lllll}
\hline
\textbf{\begin{tabular}[c]{@{}l@{}}Number\\ of \\ Sentences\end{tabular}} & \textbf{\begin{tabular}[c]{@{}l@{}}Number of \\ unique \\ characters\\ contained\end{tabular}} & \textbf{\begin{tabular}[c]{@{}l@{}}Number of\\ unique\\ words\\ contained\end{tabular}} & \textbf{\begin{tabular}[c]{@{}l@{}}CTC \\ Model:\\ Listen\\ EEG\\ (CER \%)\end{tabular}} & \textbf{\begin{tabular}[c]{@{}l@{}}Attention\\  Model:\\ Listen\\ EEG\\ (WER \%)\end{tabular}} \\ \hline
3                                                                         & {\color[HTML]{000000} 19}                                                                      & 19                                                                                      & 59                                                                                       & 0                                                                                              \\ \hline
5                                                                         & {\color[HTML]{000000} 20}                                                                      & 29                                                                                      & 65                                                                                       & 48                                                                                             \\ \hline
7                                                                         & 22                                                                                             & 42                                                                                      & 70                                                                                       & 64.4                                                                                           \\ \hline
9                                                                         & {\color[HTML]{000000} 23}                                                                      & 55                                                                                      & 73                                                                                       & 68.3                                                                                           \\ \hline
\end{tabular}
\end{table}

\begin{table}[!ht]

\caption{Results for data set B with feature set 1}

\centering
\begin{tabular}{lllll}
\hline
\textbf{\begin{tabular}[c]{@{}l@{}}Number\\ of \\ Sentences\end{tabular}} & \textbf{\begin{tabular}[c]{@{}l@{}}Number of \\ unique \\ characters\\ contained\end{tabular}} & \textbf{\begin{tabular}[c]{@{}l@{}}Number of\\ unique\\ words\\ contained\end{tabular}} & \multicolumn{1}{c}{\textbf{\begin{tabular}[c]{@{}c@{}}CTC \\ Model:\\ Listen \\ EEG \\ +\\ Spoken\\ EEG\\ (CER \%)\end{tabular}}} & \multicolumn{1}{c}{\textbf{\begin{tabular}[c]{@{}c@{}}Attention\\  Model:\\ Listen\\ EEG\\ +\\ Spoken\\ EEG\\ (WER \%)\end{tabular}}} \\ \hline
3                                                                         & {\color[HTML]{000000} 19}                                                                      & 19                                                                                      & 36                                                                                                                                & 0                                                                                                                                     \\ \hline
5                                                                         & {\color[HTML]{000000} 20}                                                                      & 29                                                                                      & 65                                                                                                                                & 54                                                                                                                                    \\ \hline
7                                                                         & 22                                                                                             & 42                                                                                      & 68                                                                                                                                & 66.6                                                                                                                                  \\ \hline
9                                                                         & {\color[HTML]{000000} 23}                                                                      & 55                                                                                      & 68                                                                                                                                & 61.6                                                                                                                                  \\ \hline
\end{tabular}
\end{table}

\begin{table}[!ht]

\caption{Results for data set B with feature set 2}

\centering

\begin{tabular}{lllll}
\hline
\textbf{\begin{tabular}[c]{@{}l@{}}Number\\ of \\ Sentences\end{tabular}} & \textbf{\begin{tabular}[c]{@{}l@{}}Number of \\ unique \\ characters\\ contained\end{tabular}} & \textbf{\begin{tabular}[c]{@{}l@{}}Number of\\ unique\\ words\\ contained\end{tabular}} & \multicolumn{1}{c}{\textbf{\begin{tabular}[c]{@{}c@{}}CTC \\ Model:\\ Spoken\\ EEG\\ (CER \%)\end{tabular}}} & \multicolumn{1}{c}{\textbf{\begin{tabular}[c]{@{}c@{}}Attention\\  Model:\\ Spoken\\ EEG\\ (WER \%)\end{tabular}}} \\ \hline
3                                                                         & {\color[HTML]{000000} 19}                                                                      & 19                                                                                      & 27.5                                                                                                         & 0                                                                                                                  \\ \hline
5                                                                         & {\color[HTML]{000000} 20}                                                                      & 29                                                                                      & 57.3                                                                                                         & 45                                                                                                                 \\ \hline
7                                                                         & 22                                                                                             & 42                                                                                      & 67                                                                                                           & 57                                                                                                                 \\ \hline
9                                                                         & {\color[HTML]{000000} 23}                                                                      & 55                                                                                      & 72                                                                                                           & 70                                                                                                                 \\ \hline
\end{tabular}
\end{table}

\begin{table}[!ht]

\caption{Results for data set B with feature set 3 (Listen)}

\centering
\begin{tabular}{lllll}
\hline
\textbf{\begin{tabular}[c]{@{}l@{}}Number\\ of \\ Sentences\end{tabular}} & \textbf{\begin{tabular}[c]{@{}l@{}}Number of \\ unique \\ characters\\ contained\end{tabular}} & \textbf{\begin{tabular}[c]{@{}l@{}}Number of\\ unique\\ words\\ contained\end{tabular}} & \multicolumn{1}{c}{\textbf{\begin{tabular}[c]{@{}c@{}}CTC \\ Model:\\ Listen\\ EEG\\ (CER \%)\end{tabular}}} & \multicolumn{1}{c}{\textbf{\begin{tabular}[c]{@{}c@{}}Attention\\  Model:\\ Listen\\ EEG\\ (WER \%)\end{tabular}}} \\ \hline
3                                                                         & {\color[HTML]{000000} 19}                                                                      & 19                                                                                      & 48.3                                                                                                         & 0                                                                                                                  \\ \hline
5                                                                         & {\color[HTML]{000000} 20}                                                                      & 29                                                                                      & 63.1                                                                                                         & 41                                                                                                                 \\ \hline
7                                                                         & 22                                                                                             & 42                                                                                      & 62                                                                                                           & 51                                                                                                                 \\ \hline
9                                                                         & {\color[HTML]{000000} 23}                                                                      & 55                                                                                      & 61                                                                                                           & 66                                                                                                                 \\ \hline
\end{tabular}
\end{table}

\begin{table}[!ht]

\caption{Results for data set B with feature set 3 (Spoken)}

\centering

\begin{tabular}{lllll}
\hline
\textbf{\begin{tabular}[c]{@{}l@{}}Number\\ of \\ Sentences\end{tabular}} & \textbf{\begin{tabular}[c]{@{}l@{}}Number of \\ unique \\ characters\\ contained\end{tabular}} & \textbf{\begin{tabular}[c]{@{}l@{}}Number of\\ unique\\ words\\ contained\end{tabular}} & \multicolumn{1}{c}{\textbf{\begin{tabular}[c]{@{}c@{}}CTC \\ Model:\\ Spoken\\ EEG\\ (CER \%)\end{tabular}}} & \multicolumn{1}{c}{\textbf{\begin{tabular}[c]{@{}c@{}}Attention\\  Model:\\ Spoken\\ EEG\\ (WER \%)\end{tabular}}} \\ \hline
3                                                                         & {\color[HTML]{000000} 19}                                                                      & 19                                                                                      & 40                                                                                                          & 0                                                                                                                 \\ \hline
5                                                                         & {\color[HTML]{000000} 20}                                                                      & 29                                                                                      & 54                                                                                                          & 32                                                                                                                \\ \hline
7                                                                         & 22                                                                                             & 42                                                                                      & 62                                                                                                          & 62                                                                                                                \\ \hline
9                                                                         & {\color[HTML]{000000} 23}                                                                      & 55                                                                                      & 63                                                                                                          & 68                                                                                                                \\ \hline
\end{tabular}
\end{table}

\begin{table}[!ht]

\caption{Results for predicting listen MFCC from listen EEG feature set 1}

\centering
\begin{tabular}{llll}
\hline
\textbf{Model}                                            & \textbf{\begin{tabular}[c]{@{}l@{}}Average\\ Normalized\\ RMSE\end{tabular}} & \multicolumn{1}{c}{\textbf{\begin{tabular}[c]{@{}c@{}}Average\\ RMSE\end{tabular}}} & \multicolumn{1}{c}{\textbf{\begin{tabular}[c]{@{}c@{}}Average\\ MCD\end{tabular}}} \\ \hline
GAN                                                       & 0.216                                                                        & 63.25                                                                               & 10.926                                                                            \\ \hline
WGAN                                                      & 0.201                                                                        & 60.45                                                                               & 10.356                                                                          \\ \hline
\begin{tabular}[c]{@{}l@{}}LSTM\\ Regression\end{tabular} & 0.0291                                                                       & 6.45                                                                                & 1.413                                                                             \\ \hline
\end{tabular}
\end{table}

\begin{table}[!ht]

\caption{Results for predicting listen MFCC from listen EEG feature set 2}

\centering
\begin{tabular}{llll}
\hline
\textbf{Model}                                            & \textbf{\begin{tabular}[c]{@{}l@{}}Average\\ Normalized\\ RMSE\end{tabular}} & \multicolumn{1}{c}{\textbf{\begin{tabular}[c]{@{}c@{}}Average\\ RMSE\end{tabular}}} & \multicolumn{1}{c}{\textbf{\begin{tabular}[c]{@{}c@{}}Average\\ MCD\end{tabular}}} \\ \hline
GAN                                                       & 0.209                                                                        & 64.018                                                                              & 10.818                                                                           \\ \hline
WGAN                                                      & 0.2                                                                          & 61.21                                                                               & 10.336                                                                           \\ \hline
\begin{tabular}[c]{@{}l@{}}LSTM\\ Regression\end{tabular} & 0.0266                                                                       & 7.954                                                                               & 1.34                                                                            \\ \hline
\end{tabular}
\end{table}

\begin{table}[!ht]

\caption{Results for predicting listen MFCC from listen EEG feature set 3}

\centering
\begin{tabular}{llll}
\hline
\textbf{Model}                                            & \textbf{\begin{tabular}[c]{@{}l@{}}Average\\ Normalized\\ RMSE\end{tabular}} & \multicolumn{1}{c}{\textbf{\begin{tabular}[c]{@{}c@{}}Average\\ RMSE\end{tabular}}} & \multicolumn{1}{c}{\textbf{\begin{tabular}[c]{@{}c@{}}Average\\ MCD\end{tabular}}} \\ \hline
GAN                                                       & 0.208                                                                        & 63.12                                                                               & 10.748                                                                           \\ \hline
WGAN                                                      & 0.209                                                                        & 63.27                                                                               & 10.578                                                                           \\ \hline
\begin{tabular}[c]{@{}l@{}}LSTM\\ Regression\end{tabular} & 0.0289                                                                       & 8.526                                                                               & 1.444                                                                            \\ \hline
\end{tabular}
\end{table}

\begin{table}[!ht]

\caption{Results for predicting spoken MFCC from spoken EEG}

\centering
\begin{tabular}{llll}
\hline
\textbf{Model}                                            & \textbf{\begin{tabular}[c]{@{}l@{}}Average\\ Normalized\\ RMSE\end{tabular}} & \multicolumn{1}{c}{\textbf{\begin{tabular}[c]{@{}c@{}}Average\\ RMSE\end{tabular}}} & \multicolumn{1}{c}{\textbf{\begin{tabular}[c]{@{}c@{}}Average\\ MCD\end{tabular}}} \\ \hline
GAN                                                       & 0.193                                                                        & 73.77                                                                               & 13.249                                                                           \\ \hline
WGAN                                                      & 0.188                                                                        & 72.149                                                                              & 12.953                                                                                   \\ \hline
\begin{tabular}[c]{@{}l@{}}LSTM\\ Regression\end{tabular} & 0.126                                                                        & 48.449                                                                              & 5.737                                                                            \\ \hline
\end{tabular}
\end{table}

\section{Discussion}

From neuroscience perspective, listen EEG can be considered as the brain activity of a subject while hearing and processing intend to speak. Intend to speak includes the brain activity of the subjects responsible for recognizing phonemes present, words present, meaning, start and end of the utterances that they were hearing. 
Spoken EEG can be considered as the brain activity of a subject while articulating the speech.  Concatenation of listen and spoken EEG can be considered as mixture of brain activity involving hearing, intend to speak and articulation.  

We believe speech recognition using listen EEG will help with speech restoration for people who can not speak at all and speech recognition using spoken EEG will help with improving quality of speech for people with speaking difficulties like broken speech.
Speech recognition using concatenation of listen and spoken EEG is an interesting area which need further exploration. We believe it will help with speech restoration for people suffering from mild to severe speaking disabilities. Our work provides only preliminary results for speech recognition using EEG.

\section{Conclusion and Future work}

In this paper we demonstrated continuous noisy speech recognition using different EEG feature sets and we demonstrated LSTM based regression method, GAN model to predict listen acoustic features from listen EEG features, to predict spoken acoustic features from spoken EEG features with very low normalized RMSE during test time. We observed that for speech recognition using EEG, we were able to achieve low error rates during test time for smaller corpus size and error rates went up as we increased the corpus size. We further observed that attention model and CTC model demonstrated better performance than RNN transducer model for speech recognition using EEG.
We observed that for predicting MFCC features from EEG features, LSTM based regression model demonstrated better test time performance than GAN and WGAN model. We believe the speech recognition results can be improved by training the models with more number of examples. Another possible reason for low speech recognition accuracy might be the nature of the data set used. Our data set had a mix of non native and native English speakers with majority of the subjects being non native English speakers for both the data bases. For future work we would like to conduct experiments with equal number of training examples from both native and non native speakers and investigate whether it will help in improving ASR test time performance. 

For the speech synthesis problem, griffin lim reconstruction \cite{griffin1984signal} algorithm can be used to convert the predicted listen mfcc, spoken mfcc features to interpretable audio.  As seen from our results, we observed high MCD values, we believe to reduce the test time MCD values, the models need to be trained with much larger data set. Also we observed that both GAN and WGAN models were difficult to train compared to the simple LSTM regression model and the training loss showed lot of fluctuations in convergence rate for both GAN and WGAN models.  Our initial hypothesis was that both GAN and WGAN should demonstrate better results than LSTM regression as the GAN model learns the loss function compared to a fixed loss function in the case of LSTM regression but we observed LSTM regression model demonstrating better test time results than both GAN and WGAN models. We believe GAN results can be improved by pre training the generator, adding regularization terms to the GAN loss function etc. This will be considered for our future work. 
In this work we provide only preliminary results for speech synthesis using EEG.   

For future work we would also like to build a much larger speech EEG corpus and train the model with more examples, include implicit language model during training, include an external language model during inference time and see if our results can be improved. We would also like to see if the results can be improved by concatenating all the three EEG feature sets. For future work, we would also like to perform EEG based speech recognition and speech synthesis experiments per each subject, in that case we would need to collect lot of speech EEG data per each subject. 
We plan to publish the speech EEG databases used in this work to help advancement of research in this area.

\subsubsection*{Acknowledgments}

We would like to thank Kerry Loader and Rezwanul Kabir from Dell, Austin, TX for donating us the GPU to train the models used in this work. The first author would like to thank Satyanarayana Vusirikala from CS department, UT Austin for his helpful discussion on generative adversarial networks. 

\bibliographystyle{neurips_2019}
\bibliography{neurips_2019}

\begin{thebibliography}{10}

\bibitem{krishna2019speech}
Krishna, G., C.~Tran, J.~Yu, et~al.
\newblock Speech recognition with no speech or with noisy speech.
\newblock In \emph{Acoustics, Speech and Signal Processing (ICASSP), 2019 IEEE
  International Conference on}. IEEE, 2019.

\bibitem{krishna20}
Krishna, G., C.~Tran, M.~Carnahan, et~al.
\newblock Advancing speech recognition with no speech or with noisy speech.
\newblock In \emph{2019 27th European Signal Processing Conference (EUSIPCO)}.
  IEEE, 2019.

\bibitem{anumanchipalli2019speech}
Anumanchipalli, G.~K., J.~Chartier, E.~F. Chang.
\newblock Speech synthesis from neural decoding of spoken sentences.
\newblock \emph{Nature}, 568(7753):493, 2019.

\bibitem{ramsey2017decoding}
Ramsey, N., E.~Salari, E.~Aarnoutse, et~al.
\newblock Decoding spoken phonemes from sensorimotor cortex with high-density
  ecog grids.
\newblock \emph{Neuroimage}, 2017.

\bibitem{zhao2015classifying}
Zhao, S., F.~Rudzicz.
\newblock Classifying phonological categories in imagined and articulated
  speech.
\newblock In \emph{2015 IEEE International Conference on Acoustics, Speech and
  Signal Processing (ICASSP)}, pages 992--996. IEEE, 2015.

\bibitem{yang1991auditory}
Yang, X., K.~Wang, S.~A. Shamma.
\newblock Auditory representations of acoustic signals.
\newblock Tech. rep., 1991.

\bibitem{mesgarani2011speech}
Mesgarani, N., S.~Shamma.
\newblock Speech processing with a cortical representation of audio.
\newblock In \emph{Acoustics, Speech and Signal Processing (ICASSP), 2011 IEEE
  International Conference on}, pages 5872--5875. IEEE, 2011.

\bibitem{hochreiter1997long}
Hochreiter, S., J.~Schmidhuber.
\newblock Long short-term memory.
\newblock \emph{Neural computation}, 9(8):1735--1780, 1997.

\bibitem{goodfellow2014generative}
Goodfellow, I., J.~Pouget-Abadie, M.~Mirza, et~al.
\newblock Generative adversarial nets.
\newblock In \emph{Advances in neural information processing systems}, pages
  2672--2680. 2014.

\bibitem{arjovsky2017wasserstein}
Arjovsky, M., S.~Chintala, L.~Bottou.
\newblock Wasserstein generative adversarial networks.
\newblock In \emph{International Conference on Machine Learning}, pages
  214--223. 2017.

\bibitem{graves2006connectionist}
Graves, A., S.~Fern{\'a}ndez, F.~Gomez, et~al.
\newblock Connectionist temporal classification: labelling unsegmented sequence
  data with recurrent neural networks.
\newblock In \emph{Proceedings of the 23rd international conference on Machine
  learning}, pages 369--376. ACM, 2006.

\bibitem{graves2014towards}
Graves, A., N.~Jaitly.
\newblock Towards end-to-end speech recognition with recurrent neural networks.
\newblock In \emph{International Conference on Machine Learning}, pages
  1764--1772. 2014.

\bibitem{cho2014learning}
Cho, K., B.~Van~Merri{\"e}nboer, C.~Gulcehre, et~al.
\newblock Learning phrase representations using rnn encoder-decoder for
  statistical machine translation.
\newblock \emph{arXiv preprint arXiv:1406.1078}, 2014.

\bibitem{chorowski2015attention}
Chorowski, J.~K., D.~Bahdanau, D.~Serdyuk, et~al.
\newblock Attention-based models for speech recognition.
\newblock In \emph{Advances in neural information processing systems}, pages
  577--585. 2015.

\bibitem{bahdanau2014neural}
Bahdanau, D., K.~Cho, Y.~Bengio.
\newblock Neural machine translation by jointly learning to align and
  translate.
\newblock \emph{arXiv preprint arXiv:1409.0473}, 2014.

\bibitem{graves2012sequence}
Graves, A.
\newblock Sequence transduction with recurrent neural networks.
\newblock \emph{arXiv preprint arXiv:1211.3711}, 2012.

\bibitem{graves2013speech}
Graves, A., A.-r. Mohamed, G.~Hinton.
\newblock Speech recognition with deep recurrent neural networks.
\newblock In \emph{2013 IEEE international conference on acoustics, speech and
  signal processing}, pages 6645--6649. IEEE, 2013.

\bibitem{chung2014empirical}
Chung, J., C.~Gulcehre, K.~Cho, et~al.
\newblock Empirical evaluation of gated recurrent neural networks on sequence
  modeling.
\newblock \emph{arXiv preprint arXiv:1412.3555}, 2014.

\bibitem{kingma2014adam}
Kingma, D.~P., J.~Ba.
\newblock Adam: A method for stochastic optimization.
\newblock \emph{arXiv preprint arXiv:1412.6980}, 2014.

\bibitem{williams1989learning}
Williams, R.~J., D.~Zipser.
\newblock A learning algorithm for continually running fully recurrent neural
  networks.
\newblock \emph{Neural computation}, 1(2):270--280, 1989.

\bibitem{narayanan2014real}
Narayanan, S., A.~Toutios, V.~Ramanarayanan, et~al.
\newblock Real-time magnetic resonance imaging and electromagnetic
  articulography database for speech production research (tc).
\newblock \emph{The Journal of the Acoustical Society of America},
  136(3):1307--1311, 2014.

\bibitem{delorme2004eeglab}
Delorme, A., S.~Makeig.
\newblock Eeglab: an open source toolbox for analysis of single-trial eeg
  dynamics including independent component analysis.
\newblock \emph{Journal of neuroscience methods}, 134(1):9--21, 2004.

\bibitem{mika1999kernel}
Mika, S., B.~Sch{\"o}lkopf, A.~J. Smola, et~al.
\newblock Kernel pca and de-noising in feature spaces.
\newblock In \emph{Advances in neural information processing systems}, pages
  536--542. 1999.

\bibitem{griffin1984signal}
Griffin, D., J.~Lim.
\newblock Signal estimation from modified short-time fourier transform.
\newblock \emph{IEEE Transactions on Acoustics, Speech, and Signal Processing},
  32(2):236--243, 1984.

\end{thebibliography}

\subsubsection*{A Additional Figures and Tables}

Tables 12, 13 and 14 shows the test time results for RNN transducer model for data set B for different experimental conditions ( Spoken, Listen, concatenation of spoken and listen). 

Table 15 shows the test time result for CTC and attention model for data set B with EEG feature set 2. 

\begin{table}[!ht]

\caption{Results for RNN Transducer model on data set B with feature set 1 (Listen)}

\centering

\begin{tabular}{lll}
\hline
\textbf{\begin{tabular}[c]{@{}l@{}}Number\\ of\\ Sentences\end{tabular}} & \textbf{\begin{tabular}[c]{@{}l@{}}Number of\\  unique\\ characters\\ contained\end{tabular}} & \textbf{\begin{tabular}[c]{@{}l@{}}Listen\\ EEG\\ (CER\%)\end{tabular}} \\ \hline
3                                                                        & {\color[HTML]{000000} 19}                                                                     & 58.91                                                                   \\ \hline
5                                                                        & {\color[HTML]{000000} 20}                                                                     & 62.03                                                                   \\ \hline
7                                                                        & 22                                                                                            & 64.43                                                                   \\ \hline
9                                                                        & 23                                                                                            & 69.30                                                                   \\ \hline
\end{tabular}
\end{table}

\begin{table}[!ht]

\caption{Results for RNN Transducer model on data set B with feature set 1 (Spoken)}

\centering

\begin{tabular}{lll}
\hline
\textbf{\begin{tabular}[c]{@{}l@{}}Number\\ of\\ Sentences\end{tabular}} & \textbf{\begin{tabular}[c]{@{}l@{}}Number of\\  unique\\ characters\\ contained\end{tabular}} & \textbf{\begin{tabular}[c]{@{}l@{}}Spoken\\ EEG\\ (CER\%)\end{tabular}} \\ \hline
3                                                                        & {\color[HTML]{000000} 19}                                                                     & 65.08                                                                   \\ \hline
5                                                                        & {\color[HTML]{000000} 20}                                                                     & 56.05                                                                   \\ \hline
7                                                                        & 22                                                                                            & 73.21                                                                   \\ \hline
9                                                                        & 23                                                                                            & 71.03                                                                   \\ \hline
\end{tabular}
\end{table}

\begin{table}[!ht]

\caption{Results for RNN Transducer model on data set B with feature set 1 (Spoken + Listen)}

\centering

\begin{tabular}{lll}
\hline
\textbf{\begin{tabular}[c]{@{}l@{}}Number\\ of\\ Sentences\end{tabular}} & \textbf{\begin{tabular}[c]{@{}l@{}}Number of\\  unique\\ characters\\ contained\end{tabular}} & \textbf{\begin{tabular}[c]{@{}l@{}}Spoken\\ EEG\\ +\\ Listen\\ EEG\\ (CER\%)\end{tabular}} \\ \hline
3                                                                        & {\color[HTML]{000000} 19}                                                                     & 65.08                                                                                      \\ \hline
5                                                                        & {\color[HTML]{000000} 20}                                                                     & 65.82                                                                                      \\ \hline
7                                                                        & 22                                                                                            & 75.45                                                                                      \\ \hline
9                                                                        & 23                                                                                            & 75.83                                                                                      \\ \hline
\end{tabular}
\end{table}

\begin{table}[!ht]

\caption{Result for Data set B with feature set 2 (Listen) }

\centering

\begin{tabular}{lllll}
\hline
\textbf{\begin{tabular}[c]{@{}l@{}}Number\\ of \\ Sentences\end{tabular}} & \textbf{\begin{tabular}[c]{@{}l@{}}Number of \\ unique \\ characters\\ contained\end{tabular}} & \textbf{\begin{tabular}[c]{@{}l@{}}Number of\\ unique\\ words\\ contained\end{tabular}} & \multicolumn{1}{c}{\textbf{\begin{tabular}[c]{@{}c@{}}CTC \\ Model:\\ Listen\\ EEG\\ (CER \%)\end{tabular}}} & \multicolumn{1}{c}{\textbf{\begin{tabular}[c]{@{}c@{}}Attention\\  Model:\\ Listen\\ EEG\\ (WER \%)\end{tabular}}} \\ \hline
3                                                                         & {\color[HTML]{000000} 19}                                                                      & 19                                                                                      & 60.5                                                                                                         & 31.5                                                                                                               \\ \hline
5                                                                         & {\color[HTML]{000000} 20}                                                                      & 29                                                                                      & 63.2                                                                                                         & 41.9                                                                                                               \\ \hline
7                                                                         & 22                                                                                             & 42                                                                                      & 71.9                                                                                                         & 57.7                                                                                                               \\ \hline
9                                                                         & {\color[HTML]{000000} 23}                                                                      & 55                                                                                      & 72.3                                                                                                         & 63.3                                                                                                               \\ \hline
\end{tabular}
\end{table}

\section{EEG sensor placement}

Figure 3 shows the EEG sensor location mapping obtained using EEG lab tool box. Figures 4 and 5 shows the three dimensional view of the EEG sensor locations. 

\begin{figure}[!ht]
\begin{center}
\includegraphics[height=3cm,width=0.25\textwidth,trim={0.1cm 0.1cm 0.1cm 0.1cm},clip]{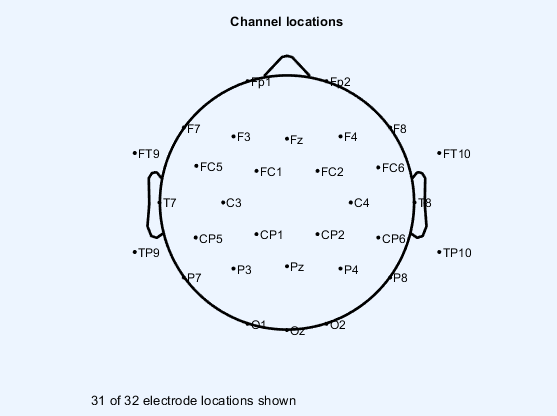}
\caption{EEG channel locations for the cap used in our experiments} 
\label{1vsall}
\end{center}
\end{figure}

\begin{figure}[!ht]
\begin{center}
\includegraphics[height=5cm,width=0.5\textwidth,trim={0.1cm 0.1cm 0.1cm 0.1cm},clip]{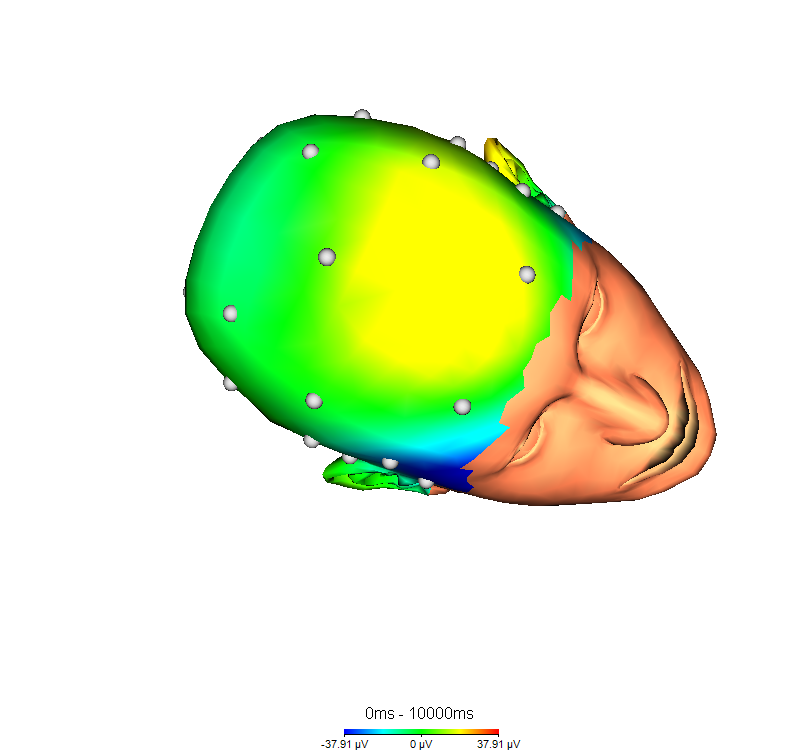}
\caption{Three dimensional view of the EEG sensor locations for the cap used in our experiments} 
\label{1vsall}
\end{center}
\end{figure}

\begin{figure}[!ht]
\begin{center}
\includegraphics[height=5cm,width=0.5\textwidth,trim={0.1cm 0.1cm 0.1cm 0.1cm},clip]{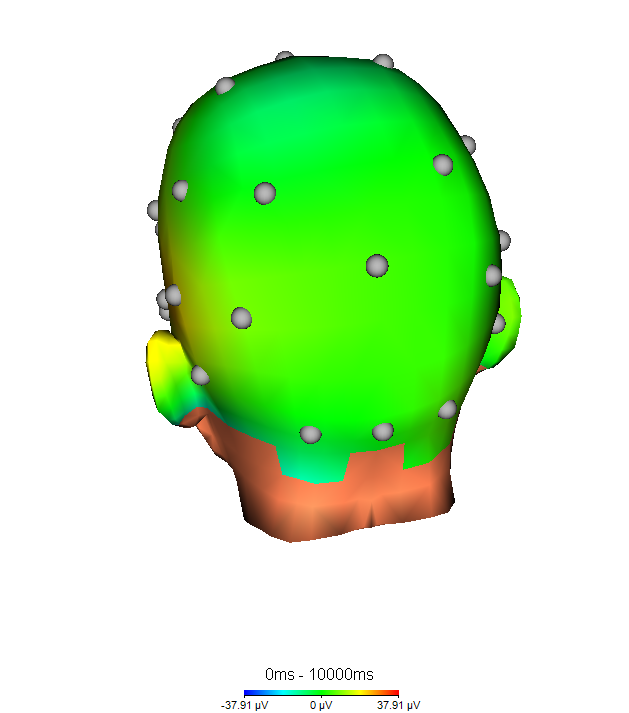}
\caption{Three dimensional view of the EEG sensor locations for the cap used in our experiments} 
\label{1vsall}
\end{center}
\end{figure}

\section{Training loss convergence plots}

Figure 6 shows the training loss convergence for attention model when trained on data set A using EEG feature set 1 for number of sentences equal to three. Figure 7 shows the training loss convergence for CTC model when trained on data set A using EEG feature set 1 for number of sentences equal to three. We observed that the training loss converged for all the models for all the experiments.

\begin{figure}[!ht]
\begin{center}
\includegraphics[height=5cm,width=0.5\textwidth,trim={0.1cm 0.1cm 0.1cm 0.1cm},clip]{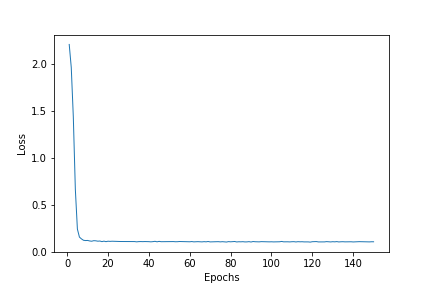}
\caption{Attention model training loss convergence} 
\label{1vsall}
\end{center}
\end{figure} 

\begin{figure}[!ht]
\begin{center}
\includegraphics[height=5cm,width=0.5\textwidth,trim={0.1cm 0.1cm 0.1cm 0.1cm},clip]{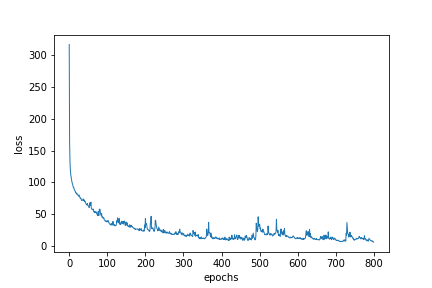}
\caption{CTC model training loss convergence} 
\label{1vsall}
\end{center}
\end{figure}

\section{Cumulative explained variance plots}

Figure 8 shows the cumulative explained variance plot for EEG feature set 1. From the plot it is clear that the optimal feature dimension required to represent the complete feature set is 30.

Figure 9 shows the cumulative explained variance plot for EEG feature set 3. From the plot it is clear that the explained variance is not converging to an optimal value, hence we didn't perform dimension reduction for this feature set. So the EEG feature dimension for feature set 3 was 93. 

Figure 10 shows the cumulative explained variance plot for EEG feature set 2. In this case it is very difficult to figure out the optimal feature dimension required to represent the complete feature set as the convergence point is not clearly observable. So we used a development set to figure out the optimal feature dimension as 50 for EEG feature set 2. 

\begin{figure}[!ht]
\begin{center}
\includegraphics[height=5cm,width=0.5\textwidth,trim={0.1cm 0.1cm 0.1cm 0.1cm},clip]{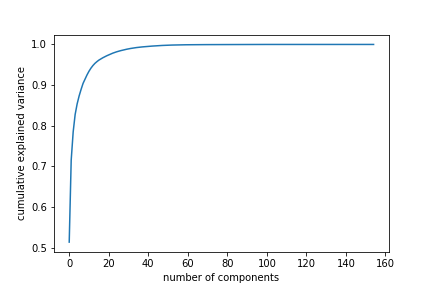}
\caption{Cumulative explained variance plot for EEG feature set 1} 
\label{1vsall}
\end{center}
\end{figure} 

\begin{figure}[!ht]
\begin{center}
\includegraphics[height=5cm,width=0.5\textwidth,trim={0.1cm 0.1cm 0.1cm 0.1cm},clip]{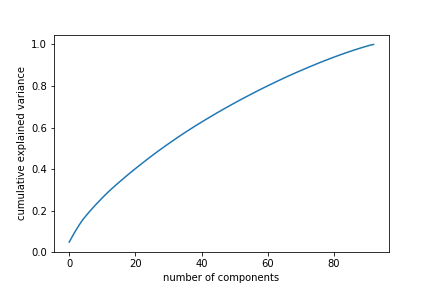}
\caption{Cumulative explained variance plot for EEG feature set 3} 
\label{1vsall}
\end{center}
\end{figure} 

\begin{figure}[!ht]
\begin{center}
\includegraphics[height=5cm,width=0.5\textwidth,trim={0.1cm 0.1cm 0.1cm 0.1cm},clip]{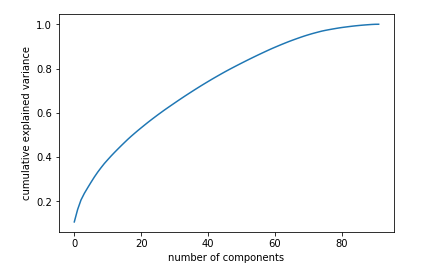}
\caption{Cumulative explained variance plot for EEG feature set 2} 
\label{1vsall}
\end{center}
\end{figure} 

\section{Additional results for predicting listen MFCC from listen EEG} 

Figures 11 and 12 shows the training loss plot for WGAN and Figures 13 and 14 shows the training loss plot for GAN for feature set 1. As seen from the figures, WGAN generator and discriminator models showed better training loss convergence than GAN. 

Figure 15 shows the training loss convergence for the LSTM regression model for feature set 1. 

Figures 16, 17 and 18 shows the test time performance of WGAN, GAN and LSTM regression models respectively for predicting listen MFCC features from listen EEG features using feature set 1. The plots shows the normalized RMSE values per each test set sample. 

Figures 19, 20 and 21 shows the test time performance of WGAN, GAN and LSTM regression models respectively for predicting listen MFCC features from listen EEG features using feature set 2.

Figures 22, 23 and 24 shows the test time performance of WGAN, GAN and LSTM regression models respectively for predicting listen MFCC features from listen EEG features using feature set 3.

Figures 25, 26 and 27 shows the test time performance of WGAN, GAN and LSTM regression models respectively for predicting spoken MFCC from spoken EEG. 


\begin{figure}[!ht]
\begin{center}
\includegraphics[height=5cm,width=0.5\textwidth,trim={0.1cm 0.1cm 0.1cm 0.1cm},clip]{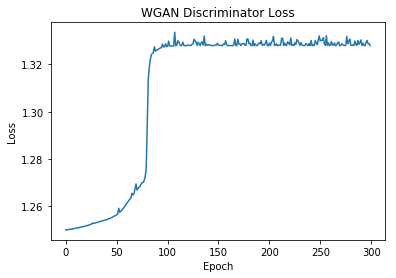}
\caption{Training loss for WGAN discriminator model} 
\label{1vsall}
\end{center}
\end{figure} 

\begin{figure}[!ht]
\begin{center}
\includegraphics[height=5cm,width=0.5\textwidth,trim={0.1cm 0.1cm 0.1cm 0.1cm},clip]{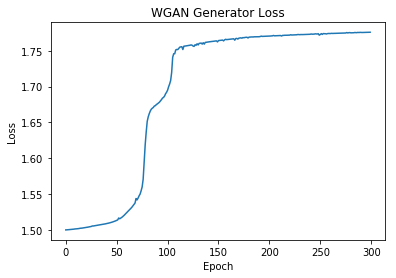}
\caption{Training loss for WGAN generator model} 
\label{1vsall}
\end{center}
\end{figure} 

\begin{figure}[!ht]
\begin{center}
\includegraphics[height=5cm,width=0.5\textwidth,trim={0.1cm 0.1cm 0.1cm 0.1cm},clip]{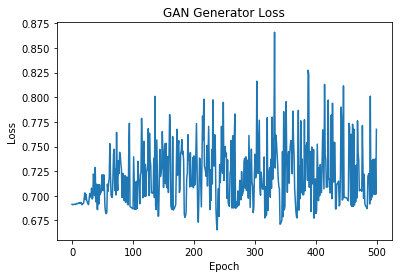}
\caption{Training loss for GAN generator model} 
\label{1vsall}
\end{center}
\end{figure} 

\begin{figure}[!ht]
\begin{center}
\includegraphics[height=5cm,width=0.5\textwidth,trim={0.1cm 0.1cm 0.1cm 0.1cm},clip]{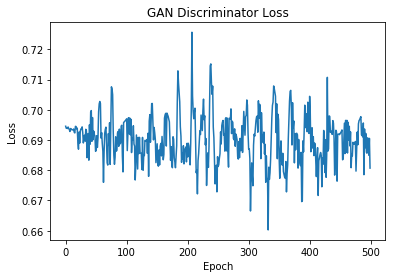}
\caption{Training loss for GAN discriminator model} 
\label{1vsall}
\end{center}
\end{figure} 

\begin{figure}[!ht]
\begin{center}
\includegraphics[height=5cm,width=0.5\textwidth,trim={0.1cm 0.1cm 0.1cm 0.1cm},clip]{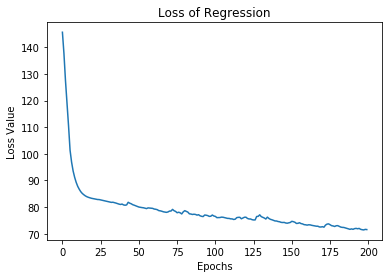}
\caption{Training loss for LSTM regression model} 
\label{1vsall}
\end{center}
\end{figure} 

\begin{figure}[!ht]
\begin{center}
\includegraphics[height=5cm,width=0.5\textwidth,trim={0.1cm 0.1cm 0.1cm 0.1cm},clip]{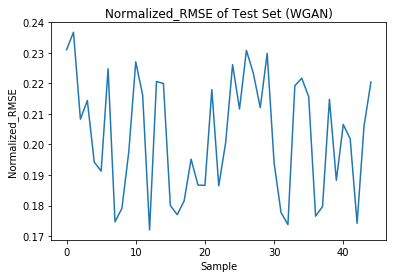}
\caption{Test time result for WGAN model with feature set 1 EEG} 
\label{1vsall}
\end{center}
\end{figure}

\begin{figure}[!ht]
\begin{center}
\includegraphics[height=5cm,width=0.5\textwidth,trim={0.1cm 0.1cm 0.1cm 0.1cm},clip]{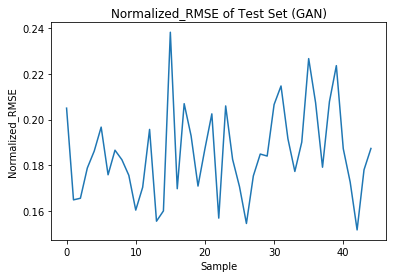}
\caption{Test time result for GAN model with feature set 1 EEG} 
\label{1vsall}
\end{center}
\end{figure}

\begin{figure}[!ht]
\begin{center}
\includegraphics[height=5cm,width=0.5\textwidth,trim={0.1cm 0.1cm 0.1cm 0.1cm},clip]{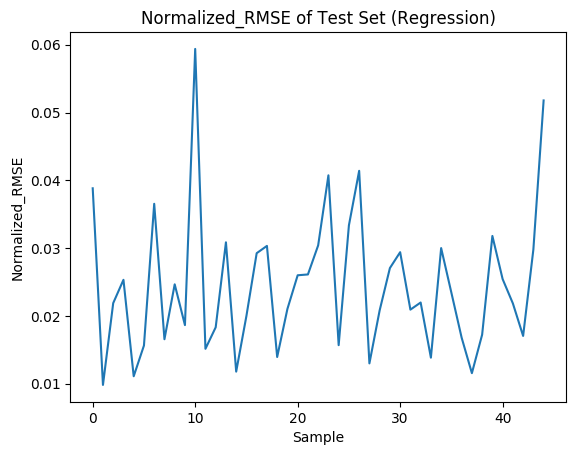}
\caption{Test time result for LSTM regression model with feature set 1 EEG} 
\label{1vsall}
\end{center}
\end{figure}

\begin{figure}[!ht]
\begin{center}
\includegraphics[height=5cm,width=0.5\textwidth,trim={0.1cm 0.1cm 0.1cm 0.1cm},clip]{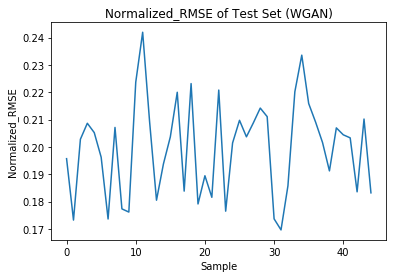}
\caption{Test time result for WGAN model with feature set 2 EEG} 
\label{1vsall}
\end{center}
\end{figure}

\begin{figure}[!ht]
\begin{center}
\includegraphics[height=5cm,width=0.5\textwidth,trim={0.1cm 0.1cm 0.1cm 0.1cm},clip]{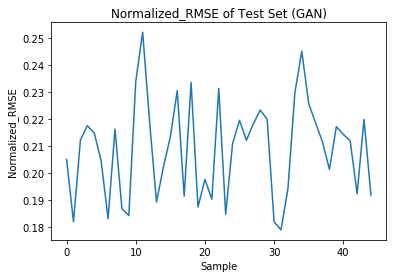}
\caption{Test time result for GAN model with feature set 2 EEG} 
\label{1vsall}
\end{center}
\end{figure}

\begin{figure}[!ht]
\begin{center}
\includegraphics[height=5cm,width=0.5\textwidth,trim={0.1cm 0.1cm 0.1cm 0.1cm},clip]{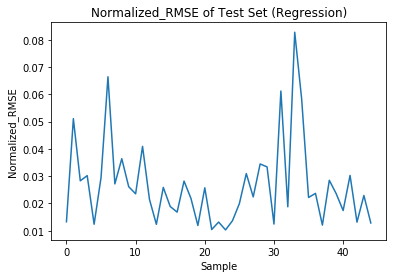}
\caption{Test time result for LSTM regression model with feature set 2 EEG} 
\label{1vsall}
\end{center}
\end{figure}

\begin{figure}[!ht]
\begin{center}
\includegraphics[height=5cm,width=0.5\textwidth,trim={0.1cm 0.1cm 0.1cm 0.1cm},clip]{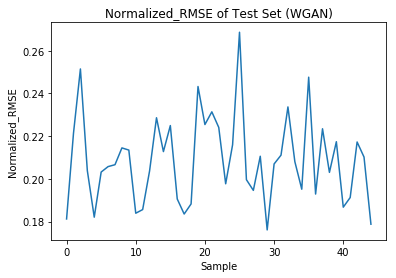}
\caption{Test time result for WGAN model with feature set 3 EEG} 
\label{1vsall}
\end{center}
\end{figure}

\begin{figure}[!ht]
\begin{center}
\includegraphics[height=5cm,width=0.5\textwidth,trim={0.1cm 0.1cm 0.1cm 0.1cm},clip]{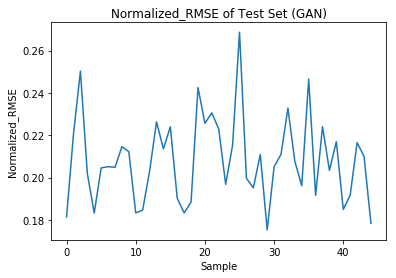}
\caption{Test time result for GAN model with feature set 3 EEG} 
\label{1vsall}
\end{center}
\end{figure}

\begin{figure}[!ht]
\begin{center}
\includegraphics[height=5cm,width=0.5\textwidth,trim={0.1cm 0.1cm 0.1cm 0.1cm},clip]{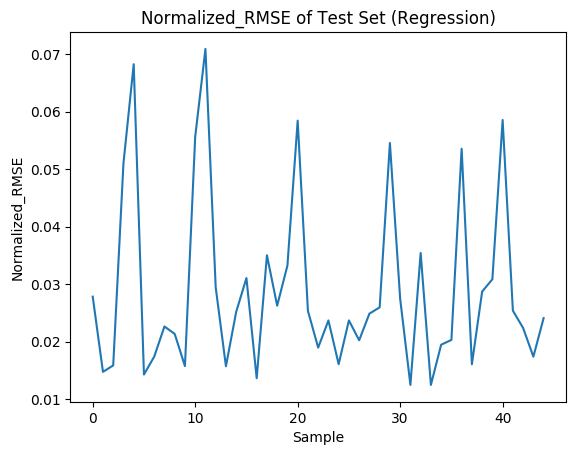}
\caption{Test time result for LSTM regression model with feature set 3 EEG} 
\label{1vsall}
\end{center}
\end{figure}

\begin{figure}[!ht]
\begin{center}
\includegraphics[height=5cm,width=0.5\textwidth,trim={0.1cm 0.1cm 0.1cm 0.1cm},clip]{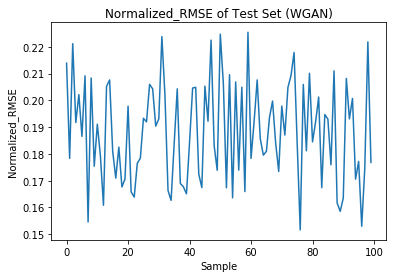}
\caption{Test time result for WGAN model for predicting spoken MFCC from spoken EEG} 
\label{1vsall}
\end{center}
\end{figure}

\begin{figure}[!ht]
\begin{center}
\includegraphics[height=5cm,width=0.5\textwidth,trim={0.1cm 0.1cm 0.1cm 0.1cm},clip]{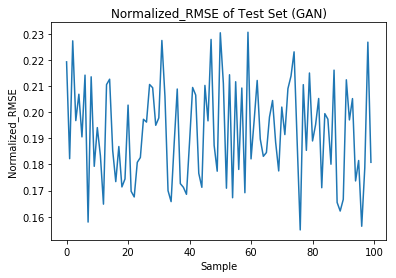}
\caption{Test time result for GAN model for predicting spoken MFCC from spoken EEG} 
\label{1vsall}
\end{center}
\end{figure}

\begin{figure}[!ht]
\begin{center}
\includegraphics[height=5cm,width=0.5\textwidth,trim={0.1cm 0.1cm 0.1cm 0.1cm},clip]{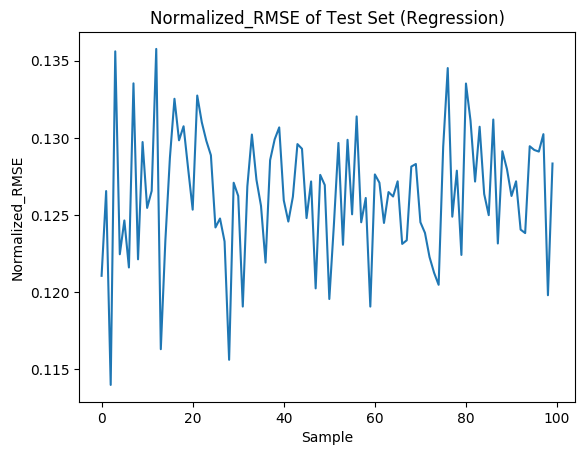}
\caption{Test time result for LSTM regression model for predicting spoken MFCC from spoken EEG} 
\label{1vsall}
\end{center}
\end{figure}


\end{document}